\documentclass[11pt]{article}
\usepackage{jcappub}
\usepackage{geometry}
\usepackage{graphicx}
\usepackage{xcolor}
\usepackage{amsmath}
\usepackage{amsfonts}
\usepackage{amssymb}
\usepackage{ascmac}
\usepackage[T1]{fontenc}
\usepackage{tikz}
\usetikzlibrary{positioning,calc}
\usepackage[compat=1.1.0]{tikz-feynhand}
\usepackage{enumerate}
\definecolor{red}{cmyk}{0,0.8,1,0}
\definecolor{blue}{cmyk}{1,0.5,0,0}
\definecolor{green}{cmyk}{0.97,0,0.75,0}
\definecolor{cyan}{cmyk}{0.8,0,0,0}
\definecolor{magenta}{cmyk}{0.1,0.7,0,0}
\definecolor{yellow}{cmyk}{0.1,0.05,0.9,0}
\definecolor{orange}{cmyk}{0,0.5,1,0}
 
\title{
    Accelerating universe at early and late times in extended Jordan-Brans-Dicke gravity
}

\author[a]{Kunio Kaneta,}
\emailAdd{kaneta@ed.niigata-u.ac.jp}
\author[b]{Kin-ya Oda,}
\emailAdd{odakin@lab.twcu.ac.jp}
\author[c]{Motohiko Yoshimura}
\emailAdd{yoshim@okayama-u.ac.jp}

\affiliation[a]{Faculty of Education, Niigata University, \\Niigata 050-2181, Japan}
\affiliation[b]{Department of Mathematics, Tokyo Woman’s Christian University, \\Tokyo 167-8585, Japan}
\affiliation[c]{Research Institute for Interdisciplinary Science, Okayama University, \\Okayama 700-8530, Japan}

\abstract{
    We propose a scenario that can explain the early-time inflation and the late-time dark energy within a unified framework.
    A scalar potential combining power-law and exponential type in a context of extended Jordan-Brans-Dicke gravity is critically important for this realization. A realistic scenario can be achieved in a two-field model in which one directional motion in field space realizes the slow-roll inflation.
    The inflaton ends up with oscillatory period and turns its direction to another direction that is identified as the quintessence field, giving rise to the dark energy at late times.
    The inflaton oscillation is expected to realize efficient heating if parametric amplification works.
    Along the quintessence direction, the present universe is on the way to reach the asymptotic fixed point.
    We search for successful parameter region, taking potential function in the form of low-order field powers times decreasing exponential in two dimensional field space.
}

\keywords{Inflation, Dark energy, Jordan-Brans-Dicke gravity}

\begin{document}
\maketitle

% \newpage
\section{Introduction}
The great success of the Standard Model (SM) of particle physics has highlighted the necessity for a new framework to comprehend the unexplained observational data in cosmology. This new physics includes dark matter~\cite{Zwicky:1933gu,1939LicOB..19...41B,Gunn:1978gr} and dark energy~\cite{SupernovaSearchTeam:1998fmf,SupernovaCosmologyProject:1998vns}, proposed to account for the universe's large-scale structure and its accelerated expansion. Additionally, cosmic inflation~\cite{Brout:1977ix,Kazanas:1980tx,Starobinsky:1980te,Sato:1981qmu,Guth:1980zm,Linde:1981mu,Albrecht:1982wi} provides a compelling framework for the initial conditions that led to structure formation~\cite{Bardeen:1980kt,Mukhanov:1981xt,Hawking:1982cz,Guth:1982ec,Starobinsky:1982ee}, supported by the cosmic microwave background (CMB) spectrum~\cite{Planck:2018jri}. However, the rapid expansion of the early universe necessitates the existence of unidentified vacuum energy. These unresolved issues underscore the need for a new framework beyond the SM, potentially involving modifications to general relativity.

An appealing approach to addressing unsolved questions in cosmology has been the framework introduced by Jordan~\cite{Jordan:1959eg} and subsequently developed by Brans and Dicke~\cite{Brans:1961sx} (JBD), which seeks a better understanding of gravity, coupled to matter fields. An extended version incorporating a potential term for the scalar field was discussed by Bergmann~\cite{Bergmann:1968ve} and Wagoner~\cite{Wagoner:1970vr}. Depending on the choice of gravitationally induced couplings of scalar fields to the SM sector, various extended models of JBD's scalar-tensor gravity can be constructed. This class of models, referred to as extended JBD (eJBD) theory, features a scalar degree of freedom that couples to gauge-invariant pieces of the Lagrangian~\cite{Kaneta:2023rby}.

The eJBD framework shares a common spirit with other models that incorporate scalar fields, such as dilaton and moduli fields. For instance, starting with a dilatation invariant theory, the spontaneous breaking of global dilatation symmetry gives rise to a Nambu-Goldstone boson, known as the dilaton. Since interacting theories often lack invariance under scaling transformations, the dilaton can couple to the trace of the energy-momentum tensor, a phenomenon known as the trace (or dilatation) anomaly~\cite{Callan:1970ze,Coleman:1970je,Coleman:1985rnk}. In the context of string theory, specific couplings between the dilaton and matter fields have been explored in Refs.~\cite{Damour:1994zq,Damour:1994ya}.
Although we do not discuss ultraviolet completion of the eJBD theory, in the present paper, we implicitly assume that an effective theory in the Einstein-metric frame emerges from the theory defined in the Jordan-metric frame where the eJBD scalars couple to all the gauge-invariant terms of the SM Lagrangian.

The properties of the eJBD scalar field can be classified into two types based on its dynamics. We refer to it as type I when the eJBD field eventually settles at a finite field value, yielding a constant vacuum energy, akin to a $\Lambda$CDM model. However, this model leaves behind the fine-tuning problem of the cosmological constant. This class includes the chameleon scenario~\cite{Khoury:2003aq,Brax:2004qh,Mota:2006fz,Wang:2012kj}, where the minimum of the potential depends on the ambient matter density due to couplings to matter fields with gravitational strength.

The other class, called type II, is characterized by a sliding behavior of the scalar eJBD field towards infinity. An example of this scenario includes quintessence models~\cite{Ratra:1987rm,Wetterich:1987fm}. If such a field is the source of the energy density driving the current expansion of the universe, the equation of state parameter $w$ can potentially deviate from $-1$, prompting more precise measurements in future observations. Furthermore, through its couplings to SM fields, physical constants may vary over time, subject to various laboratory, astrophysical, and cosmological observational constraints. A systematic study of this kind has been conducted in Ref.~\cite{Kaneta:2023rby}.

If there are more than two eJBD fields, as will be explained later, their dynamics become more intricate than in the single-field case. In this paper, we demonstrate that a two-eJBD field model provides an interesting and unified framework that can explain both inflation and dark energy on an equal footing.
In the single-field case, there have been attempts to identify the inflaton as the quintessence field that simultaneously explains dark energy~\cite{Peebles:1998qn}. A notable feature of such a scenario is that the scalar field potential does not have a global minimum at a finite field value, thus inflaton oscillation does not imply the end of inflation.\footnote{In such a non-oscillatory scenario, the gravitational production of particles~\cite{Parker:1968mv} may serve as a heating mechanism (often referred to as reheating in the literature), which is, however, inefficient, resulting in low heating temperatures. See, e.g., Ref.~\cite{Felder:1998vq} for possible remedies.}
As we will discuss, in the two-field case, inflation and dark energy can be smoothly connected during inflaton oscillation, which is expected to achieve an efficient heating mechanism.

This paper is organized as follows. 
In Sec.~\ref{sec: Scalar sector in extended Jordan-Brans-Dicke gravity}, we describe our setup based on the eJBD framework, focusing on the scalar sector.
The inflationary epoch is discussed in Sec.~\ref{sec: early time cosmology: inflation}, where we highlight that our potential parameters are strongly constrained by the spectral tilt and the tensor-to-scalar ratio.
In Sec.~\ref{sec: late time cosmology: dark energy}, we argue that the same potential shape can suitably account for dark energy in the present universe.
A two-field model connecting inflation and dark energy is proposed in Sec.~\ref{sec: Two field model connecting inflation and dark energy}, and Sec.~\ref{sec: Summary and discussion} is devoted to summary and discussion.

\section{Scalar sector in extended Jordan-Brans-Dicke gravity}
\label{sec: Scalar sector in extended Jordan-Brans-Dicke gravity}

We begin by writing the scalar-tensor sector of the Lagrangian as
\begin{align}
    \sqrt{-g_{\rm J}}{\cal L}
    &=
    \sqrt{-g_{\rm J}}\left[
        -\frac{M_\text{P}^2}{2}F_g(\phi)R_{\rm J} + \frac{1}{2}F_{{\rm d}\phi}(\phi)g_{\rm J}^{\mu\nu}\partial_\mu\phi\partial_\nu\phi - V(\phi)
    \right],
\end{align}
with $M_\text{P}\simeq 2.4\times10^{18}$ GeV being the reduced Planck mass and $F_g(\phi)$ and $F_{{\rm d}\phi}(\phi)$ being arbitrary functions of a real scalar $\phi$ coupled to the Ricci scalar $R_{\rm J}$ and metric $g_{\rm J}^{\mu\nu}$, where the subscript J refers to functions defined in the Jordan frame.

Through the Weyl rescaling $g_{{\rm J}\mu\nu}=F(\phi)g_{{\rm E}\mu\nu}$, we obtain the action in the Einstein frame,
\begin{align}
    S
    &=
    \int d^4x\sqrt{-g_{\rm E}}
    \left[
        -\frac{M_\text{P}^2}{2}R_{\rm E} + \frac{1}{2}g_{\rm E}^{\mu\nu}\partial_\mu\chi\partial_\nu\chi - V_{\rm eff}(\chi)
    \right],
\end{align}
where $\chi$ is a canonically normalized scalar field~\cite{Kaneta:2023rby}, appropriately defined in terms of $F_g(\phi)$ and $F_{{\rm d}\phi}(\phi)$.
The scalar potential $V(\chi)$ is related with $F_g(\phi)$ as
\begin{align}
    V_{\rm eff}(\chi) &= \frac{V(\phi)}{F_g(\phi)}
\end{align}
being subject to the field redefinition, $\phi=\phi(\chi)$, in the right hand side of the equation.
In the rest of the paper, we drop the subscript `eff'.

Among various possibilities for $V(\chi)$, we consider a scalar potential that combines power-law and exponential type,\footnote{
For a possible origin of such potential, see Appendix~\ref{appendix: Linear-times-exponential potential}.
}
parametrized by
\begin{align}
    V(\chi)
    &=
    V_0 \left(\frac{|\chi|}{M_\text{P}}\right)^p e^{-\gamma \chi/M_\text{P}},
    \label{eq: V}
\end{align}
where $V_0$, $p$, and $\gamma$ are assumed to be positive constants. Note that $p$ is allowed to be not only a positive integer but also a rational number. 
In the following, we use Planck units, $M_\text{P} = 1$, unless otherwise stated.

\section{Early time cosmology: inflation}
\label{sec: early time cosmology: inflation}

\begin{figure}%[htbp]
    \centering
    \includegraphics[width=.8\textwidth]{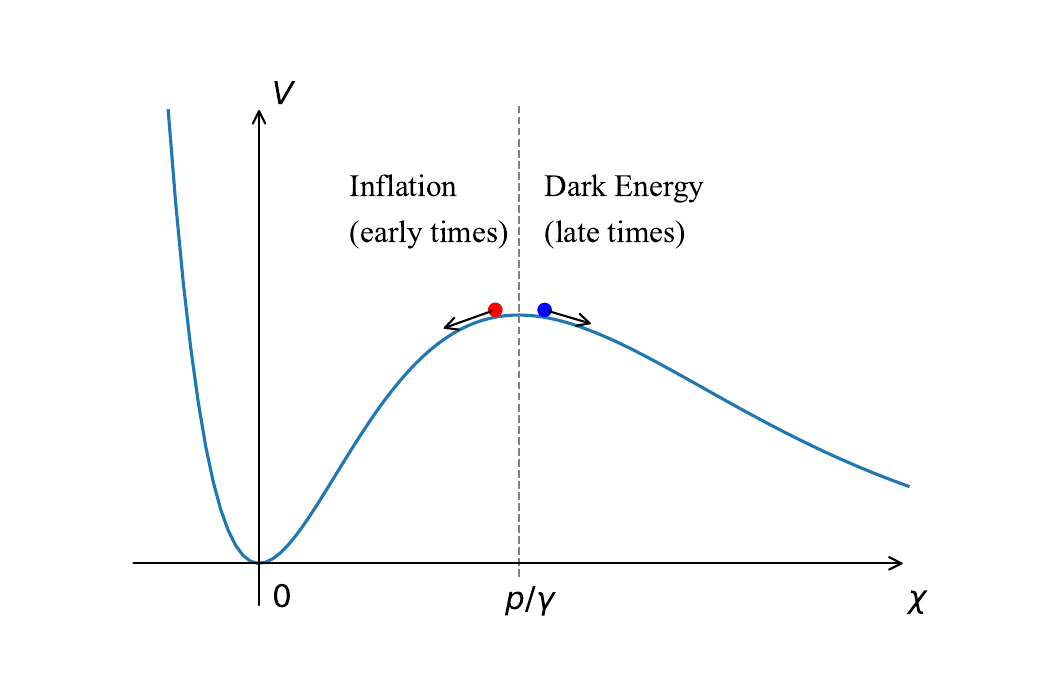}
    \caption{
    Typical potential shape of the eJBD field $\chi$. Inflation at early times is realized by slowly rolling from the potential hill toward the origin. Dark energy at late times is achieved by the field's excursion from the potential hill toward $\chi \to \infty$, where the exponential decrease dominates. In this example, we take $p=2$ and $\gamma=1$ for illustration.
    }
    \label{fig: V}
\end{figure}
In this section, we show that the potential given in Eq.~(\ref{eq: V}) explains inflation in the early universe.
The potential has a local minimum at $\chi=0$ and a local maximum at $\chi=\chi_{\rm max}=p/\gamma$. The potential exponentially grows as $\chi \to -\infty$ while exhibiting a sliding behavior as $\chi \to \infty$. Among a few possible inflation scenarios with the same potential, we particularly focus on slow-roll inflation occurring during the field's evolution from $\chi \sim \chi_{\rm max}=p/\gamma$ to $\chi=0$, as illustrated in Fig.~\ref{fig: V}. As we will discuss later, this scenario is capable of achieving reheating through the decay of the oscillating inflaton into Standard Model particles.

The dynamics of the inflaton during inflation follow the equation of motion
\begin{align}
    \ddot\chi+3H\dot\chi+\partial_\chi V = 0,
\end{align}
where the dot represents a derivative with respect to the comoving time $t$ in the Planck unit, and $H=\dot a/a$ with $a$ being the scale factor.\footnote{
We assume that a friction term caused by inflaton decay during inflation is negligible compared to the Hubble-drag term.
}
We can assess whether the slow-roll condition can be achieved by examining the slow-roll parameters~\cite{Weinberg:2008zzc} given by
\begin{align}
    \epsilon
    &\equiv
    \frac{1}{2}\left(
        \frac{\partial_\chi V}{V}
    \right)^2
    =
    \frac{1}{2}\left(
        \frac{p}{\chi}-\gamma
    \right)^2,\\
    \eta
    &\equiv 
    \frac{\partial_\chi^2V}{V}
    =
    \gamma^2-\frac{2p\gamma}{\chi}+\frac{p(p-1)}{\chi^2},
\end{align}
in the Planck unit.
Note that these quantities are independent of the overall scale of potential $V_0$.
They can indeed be sufficiently small at the vicinity of $\chi=\chi_{\rm max}$.
This scenario may be regarded as a variant of the hilltop inflation~\cite{Boubekeur:2005zm}.
At $\chi=\chi_{\rm max}$, we obtain
\begin{align}
    \epsilon(\chi=\chi_{\rm max}) 
    &= 0,
    &
    \eta(\chi=\chi_{\rm max})
    &=
    -\frac{\gamma^2}{p},
\end{align}
suggesting that $\gamma^2/p \ll 1$ should be satisfied to achieve viable inflation with $\epsilon, |\eta| \ll 1$.
The number of e-folds is defined by
\begin{align}
    N_{\rm CMB}
    &=
    \int_{\chi_{\rm end}}^{\chi_{\rm CMB}} \frac{d\chi}{\sqrt{2\epsilon}},
\end{align}
where $\chi_{\rm CMB}$ is such that the CMB scale exits the horizon at $N = N_{\rm CMB}$. 
For the definition of $\chi_{\rm end}$, see Appendix~\ref{appendix: Slow-roll conditions and the end of inflation}.

\begin{figure}%[htbp]
    \centering
    \includegraphics[width=.48\textwidth]{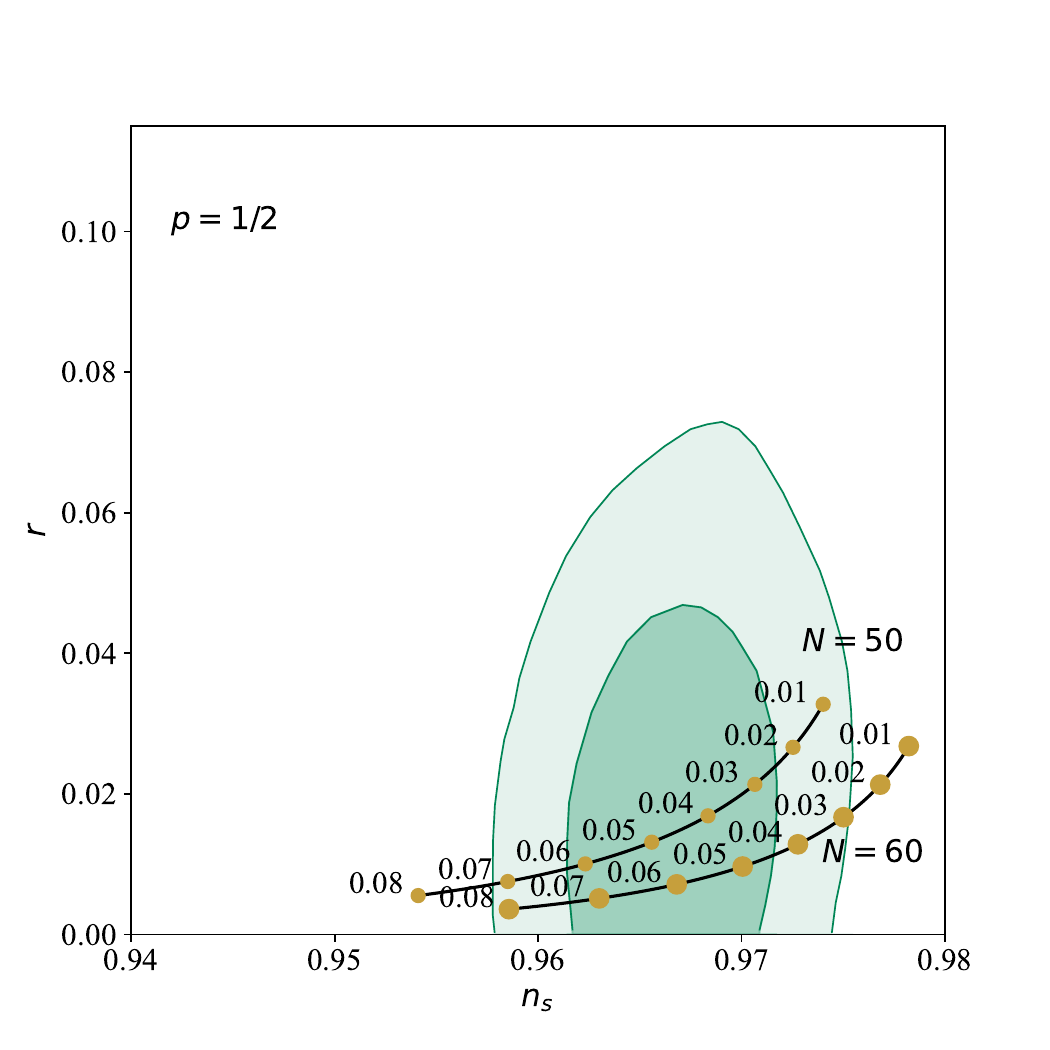}
    \includegraphics[width=.48\textwidth]{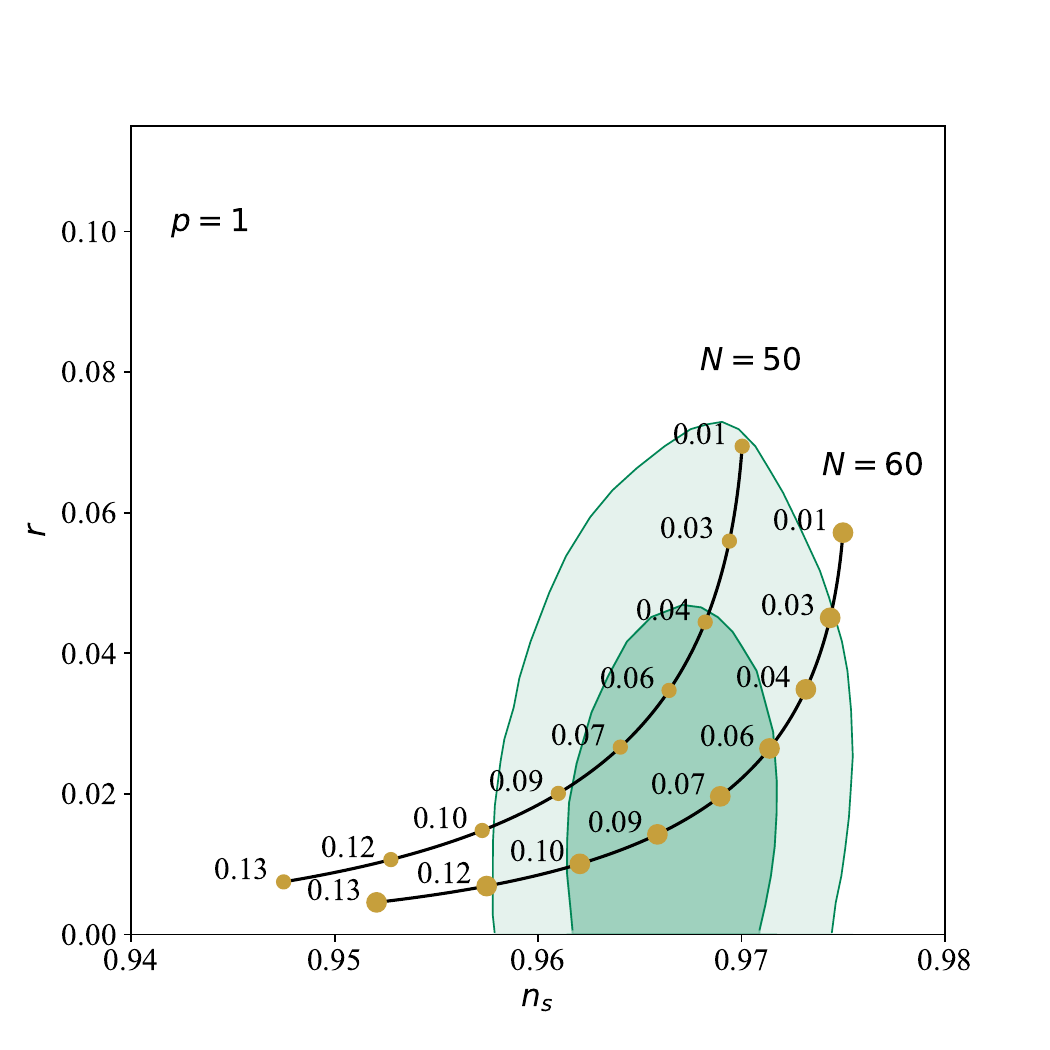}\\
    \includegraphics[width=.48\textwidth]{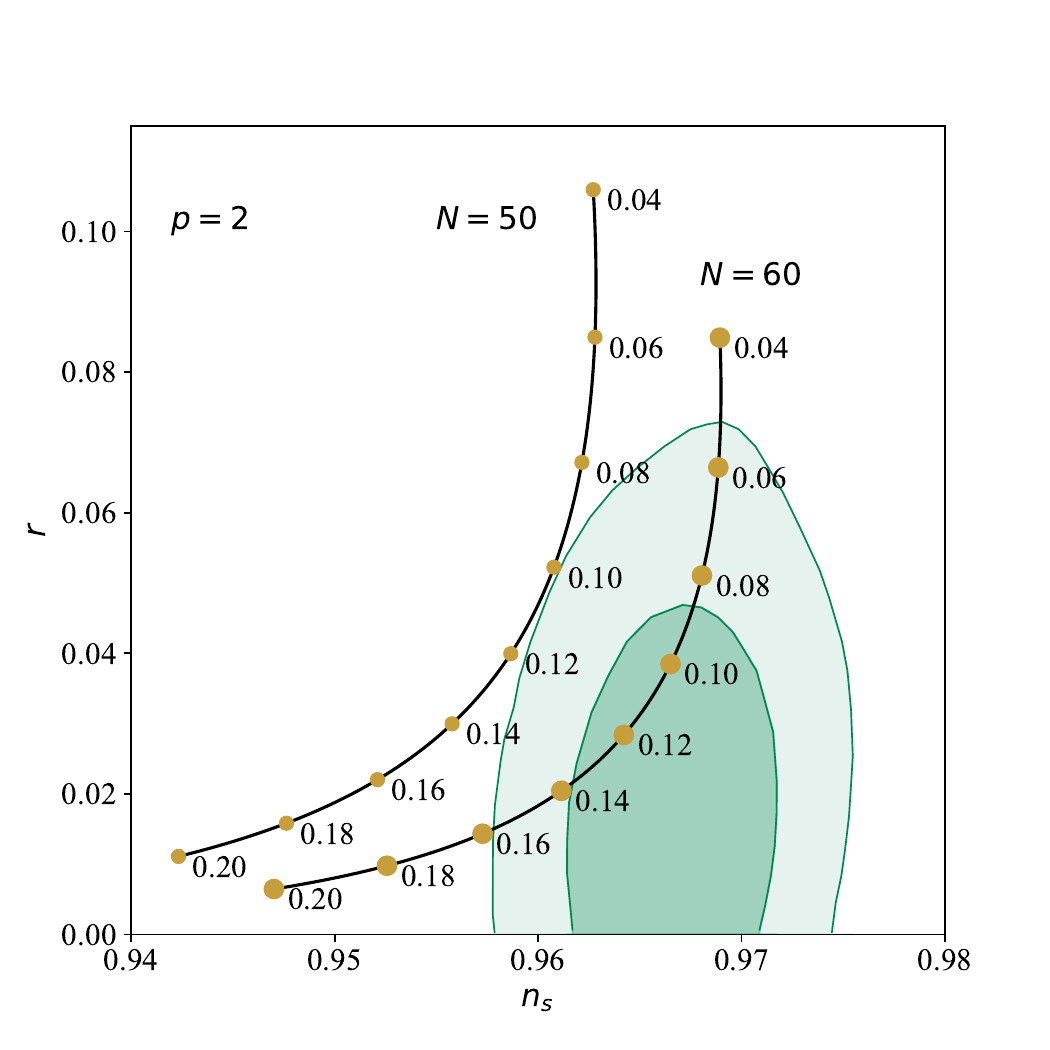}
    \includegraphics[width=.48\textwidth]{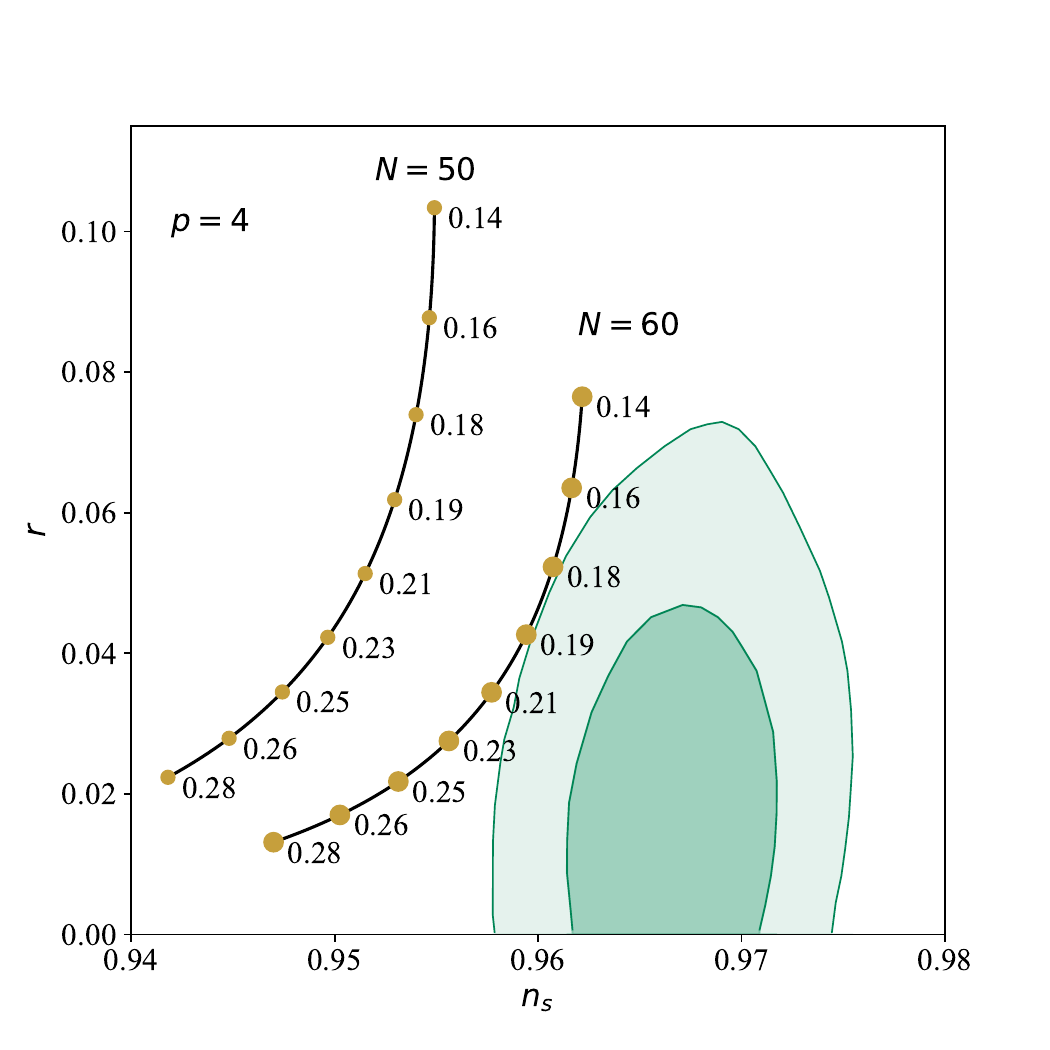}
    \caption{Inflationary predictions for $p=1/2, 1, 2, 4$. Each dot indicates the value of $\gamma$. The thick and thin hatched regions depict 68\% CL and 96\%CL area from Planck data~\cite{Planck:2018jri}, respectively.}
    \label{fig: ns-r}
\end{figure}

In Fig.~\ref{fig: ns-r}, we show the spectral index $n_s=1-6\epsilon(N_{\rm CMB})+2\eta(N_{\rm CMB})$ and the tensor-to-scalar ratio $r=16\epsilon(N_{\rm CMB})$ for the cases with $p=1/2, 1, 2, 4$, where each dot is associated with a specific choice of $\gamma$.
In all cases, small $\gamma$ values are favored to obtain a scale-invariant spectrum (close to $n_s=1$), while small $r$ is realized only when $p$ is small, $<{\cal O}(\text{a few})$.
We find that $p<4$ is needed to satisfy the Planck data~\cite{Planck:2018jri}.
The overall energy scale $V_0$ is determined by the amplitude of the scalar power spectrum $A_S\simeq 2.099\times 10^{-9}$~\cite{Planck:2018jri} via
\begin{align}
    A_S &= \frac{V(\chi_{\rm CMB})}{24\pi^2\epsilon(N_{\rm CMB})},
\end{align}
which yields $V_0^{1/4}\sim {\cal O}(10^{15}-10^{16})$ GeV, depending on $p$ and $\gamma$.

\section{Late time cosmology: dark energy}
\label{sec: late time cosmology: dark energy}

\begin{figure}%[htbp]
    \centering
    \includegraphics[width=.7\textwidth]{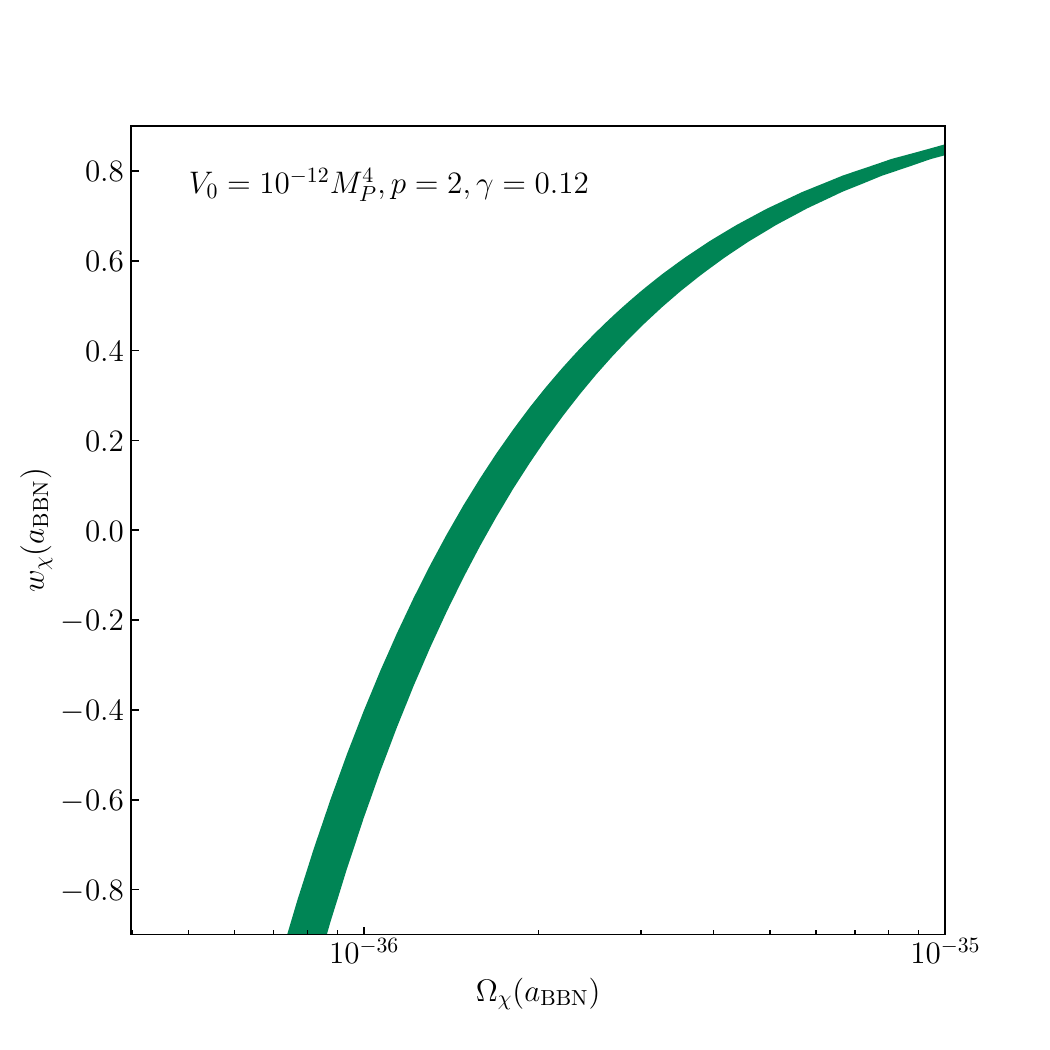}
    \caption{The colored region satisfies $\Omega_\chi(a_0)=0.6889\pm0.0168~(3\sigma)$ \cite{Planck:2018vyg}.}
    \label{fig: DE1}
\end{figure}

\begin{figure}[htbp]
    \centering
    \includegraphics[width=.7\textwidth]{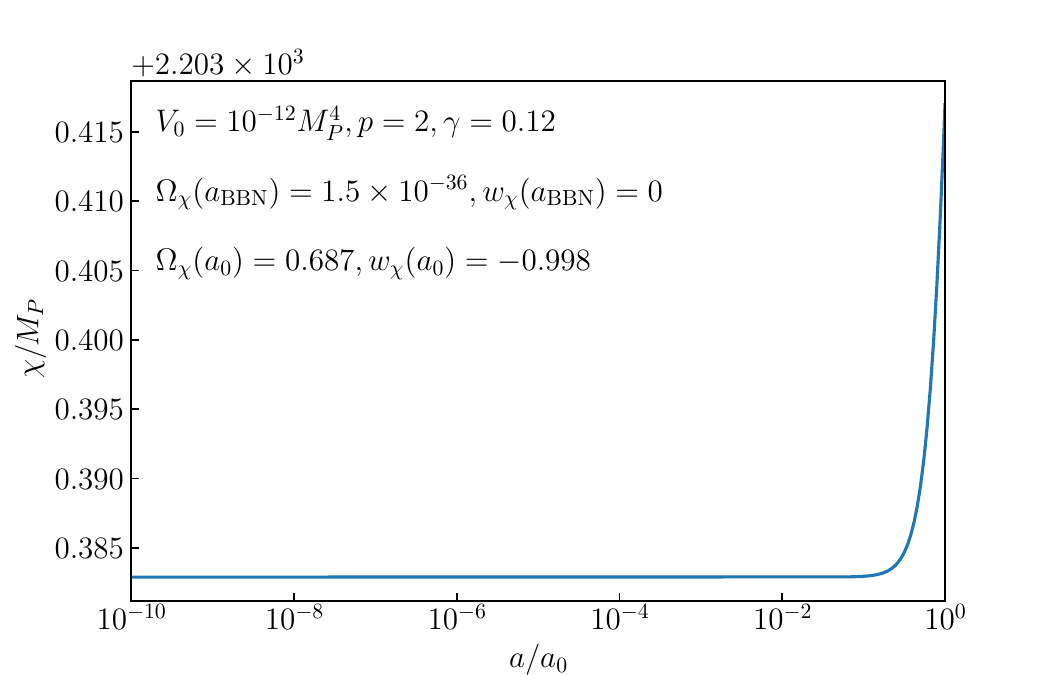}\\
    \includegraphics[width=.7\textwidth]{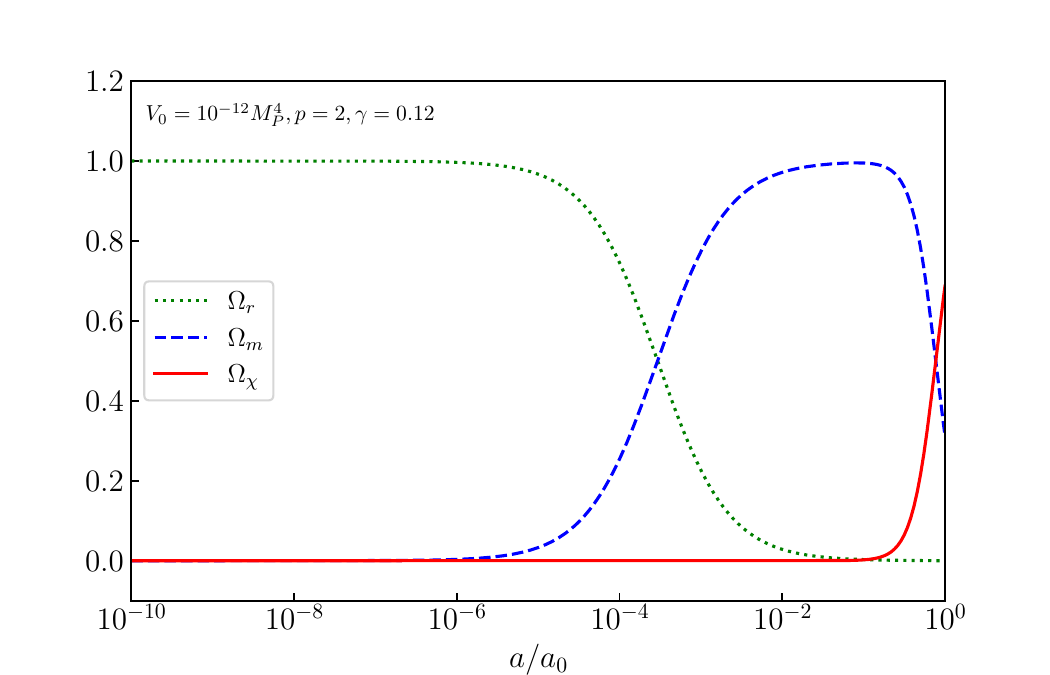}
    \caption{
    Top: Shown is a typical field evolution for a choice of viable parameters. Bottom: The evolution of the density parameters.
    }
    \label{fig: DE2}
\end{figure}

The same shape of potential (\ref{eq: V}) can also explain the present value of dark energy when $\chi\gg\chi_{\rm max}$.
In this scenario, the density parameter of the dark energy $\Omega_\chi\equiv \rho_\chi/\rho_{\rm cr}$ and the equation of state parameter $w_\chi\equiv p_\chi/\rho_\chi$ are explained by the sliding eJBD scalar, where $\rho_\chi$ and $p_\chi$ are the energy density and the pressure of $\chi$.
The critical energy density is given by $\rho_{\rm cr}\equiv 3H^2M_\text{P}^2$ whose present-day value is $\rho_{\rm cr,0}=(2.46~{\rm meV})^4$.
By writing
\begin{align}
    v
    &\equiv
    \frac{d\chi}{dN}
\end{align}
with $N\equiv \ln(a/a_0)$, we have the relations
\begin{align}
    \rho_\chi 
    &=
    K + V,\\
    w_\chi 
    &=
    \frac{K-V}{K+V},
\end{align}
where
\begin{align}
    K
    &=
    \frac{1}{2}(Hv)^2.
\end{align}
From these, we obtain
\begin{align}
    V(\chi) &= \frac{1-w_\chi}{2}\rho_\chi,\\
    v^2 &= 3(1+w_\chi)\Omega_\chi,
\end{align}
which allow us to set the boundary conditions for $\chi$ and $v$ at a given time in terms of $\Omega_\chi$ and $w_\chi$.
In our numerical analysis, we set the boundary conditions at $a/a_0 = 10^{-10}$, where $a_0$ is the scale factor at the present time.
In Fig.~\ref{fig: DE1}, the colored region satisfies $\Omega_{\chi}=0.6889\pm 0.0168~(3\sigma)$.
The parameter space shown in the figure satisfies $w_\chi(a_0)<-0.95$~\cite{Planck:2018vyg}.
The top panel of Fig.~\ref{fig: DE2} shows a typical field evolution over time, where $\chi$ is frozen until recently due to the Hubble drag force.
The bottom panel of the figure depicts the evolution of the density parameters for the same choice of parameters used in the top panel.

It is worth mentioning that the model under consideration has fixed-point solutions at $t\to\infty$.
The detailed analysis is given in Appendix~\ref{appendix: Fixed points}.
Among mathematically allowed fixed points, it turns out that the cosmologically favored fixed point is given by
\begin{align}
    \frac{\dot\chi}{\sqrt{6}H} &= \frac{\gamma}{\sqrt{6}},
    &
    \frac{\sqrt{V}}{\sqrt{3}H} &= \sqrt{1-\frac{\gamma^2}{6}},
    &
    w_\chi &= -1 + \frac{\gamma^2}{3},
\end{align}
which is an attractor solution as discussed in Appendix~\ref{appendix: Fixed points} and indicates the following two features.
First, the model predicts that the universe is completely dominated by $\chi$ at distant future, namely, $\Omega_\chi = 1$.
Therefore, what we find numerically in Figs.~\ref{fig: DE1} and \ref{fig: DE2} are the solution on the way to the fixed point, but the universe has not reached the fixed point as yet.
Second, the asymptotic fixed point does not depend on the power $p$ that appears in the potential $V$.
Thus, regardless of the choice of $p$, the dark energy can be explained if $\gamma$ is sufficiently small to achieve $w_\chi \simeq -1$.

\section{Two field model connecting inflation and dark energy}
\label{sec: Two field model connecting inflation and dark energy}

We present a model to realize both the early-time inflation and the late-time dark energy based on a new two-field framework.
Letting $\chi_1$ and $\chi_2$ denote eJBD scalar fields, we consider a potential in the Einstein frame,
\begin{align}
    V
    &=
    \left(
        \lambda_0 + m_1^2\chi_1^2 + m_2^2\chi_2^2 + \lambda_{12} \chi_1^2\chi_2^2
    \right)e^{-\gamma_1\chi_1 - \gamma_2\chi_2},
\end{align}
where $\lambda_0$, $m_1$, $m_2$, $\gamma_1$, $\gamma_2$, and $\lambda_{12}$ are positive constants.
Depending on the choice of the parameters and the initial conditions, either of the two fields plays the role of inflaton or dark energy field.
For instance, if we set the initial condition $\chi_2=\chi_{2,{\rm ini}}$ $(>0)$, $\chi_1=\dot\chi_{1,2}=0$, $\chi_2$ may be regarded as inflaton.
There are only two possibilities at infinite time: either the absolute value $\sqrt{\chi_1^2+\chi_2^2}$ goes to $\infty$ or falls into the local minimum around the field origin, with $\chi_1, \chi_2 \simeq 0$.
In the latter case, the dark energy is not realized, hence we search for the case of the sliding to field infinity.
It is obvious that the cosmological constant $\lambda_0$ presumably in the Jordan frame does not need to be fine-tuned: due to the exponential factor in $V$ the present value of the cosmological constant is nearly zero.

To illustrate the field trajectory in the model, we solve the classical equations of motion with the friction term arising from the expansion of the universe.\footnote{
Since particle production caused by inflaton oscillation is not included in our analysis, a radiation-dominated universe is not realized in this model.
}
Setting the scale factor at the initial time $a=a_i$, we introduce the number of e-folds $N=\ln(a/a_i)$.
This change of time variable is always allowed provided that $a(t)$ is monotonically increasing with time $t$.
Using the notation $'\equiv d/dN$, the equation of motion becomes
\begin{align}
    &
    \chi_i''+\frac{\dot{H}+3H^2}{H^2}\chi_i' + \frac{V_i}{H^2} = 0,
\end{align}
where $i=1,2$, $V_i\equiv \partial V/\partial\chi_i$, and
\begin{align}
    \dot{H} 
    &=
    -\frac{1}{2}H^2v^2 
\end{align}
with $v^2\equiv \chi'_1{}^2+\chi'_2{}^2$.
Assuming the negligible amount of radiation, we may approximate
\begin{align}
    3H^2 &\simeq
    \rho_{\chi_1} + \rho_{\chi_2}
    =
    \frac{V}{1-v^2/6}.
\end{align}

Recasting the potential into the form
\begin{align}
    V
    &=
    V_0 (\lambda + h \chi_1^2 + \chi_2^2 + g\chi_1^2\chi_2^2)e^{-\gamma_1\chi_1 - \gamma_2\chi_2},
\end{align}
we take $\lambda$, $h$, $g$, $\gamma_1$, and $\gamma_2$ as free parameters.
Note that $V_0$ should have been normalized properly to explain the scalar amplitude $A_S$ inferred by the Planck data; however, in the following analysis, $V_0$ can always be normalized by the timescale, namely, the choice of $H$, and hence we may take $V_0=1$ without loss of generality.
In our numerical analysis, we set $\lambda=0.1$, $h=10^{-3}$, $g=1$, $\gamma_1=0.2$, and $\gamma_2=0.12$ as an example.
The initial condition is taken as $\chi_2=11.14$, and vanishing otherwise, $\chi_1=0$, $\dot{\chi_1}=\dot{\chi_2}=0$.
This choice of $\chi_2$ respects the inflaton field value at the end of inflation for $p=2$ and $N=60$ with $\gamma = 0.12$, being consistent with the Planck data, as discussed in Sec.~\ref{sec: early time cosmology: inflation}.

\begin{figure}%[htbp]
    \centering
    \includegraphics[width=.8\textwidth]{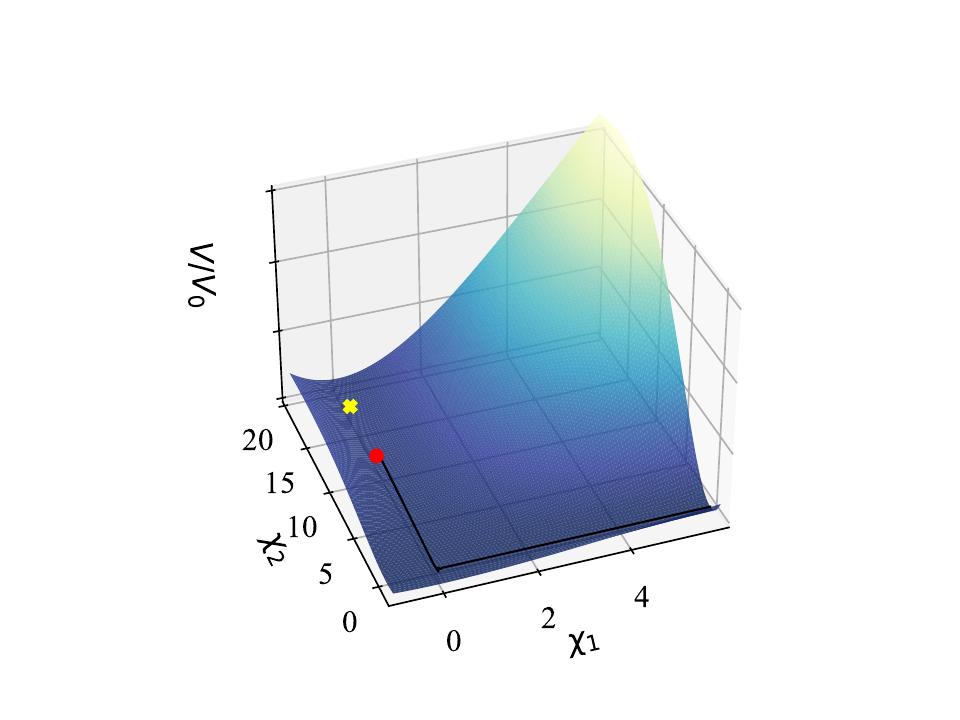}\\
    \includegraphics[width=.5\textwidth]{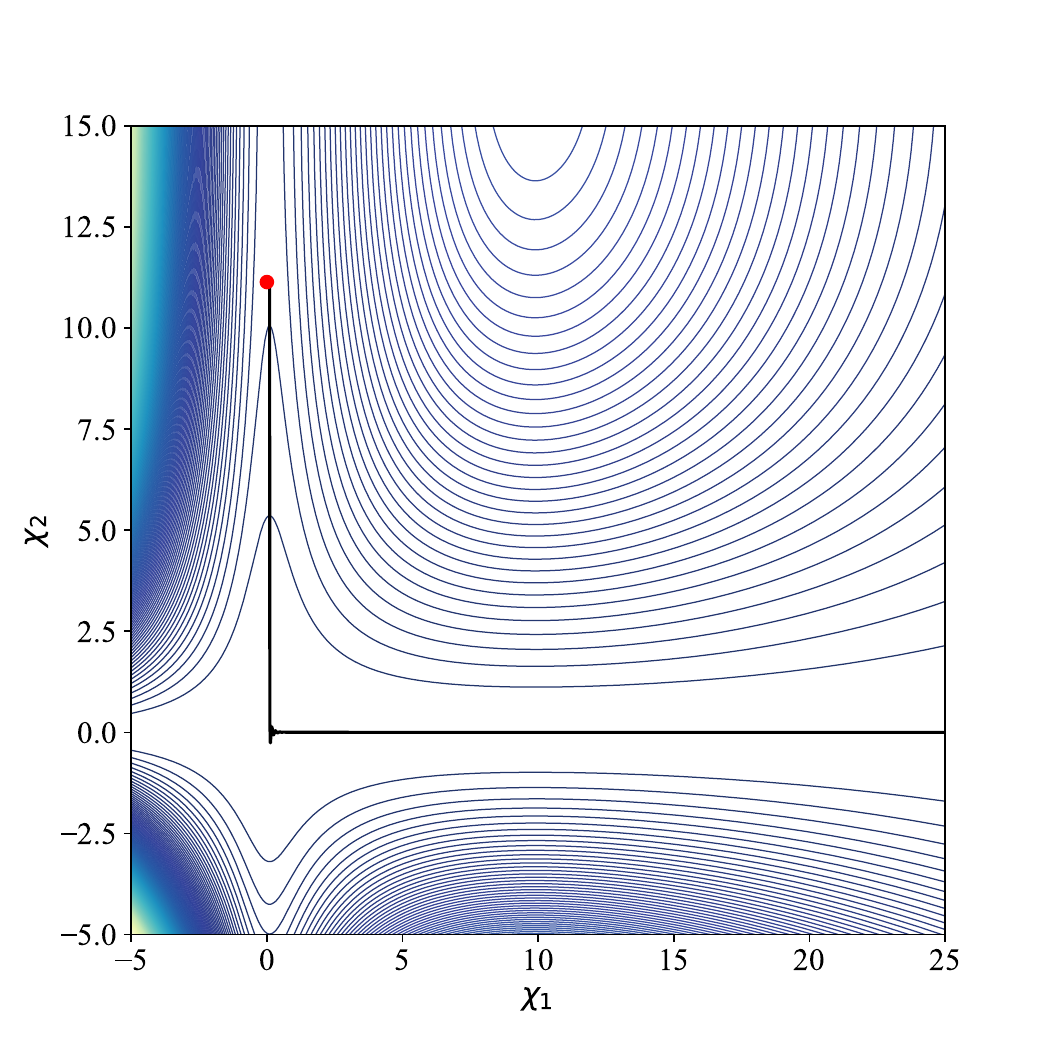}
    \caption{The top panel illustrates the trajectory in the two-field model along with the potential shape, where the black solid line is the path of the classical motion, whose starting point ($\chi_2=11.14$) is indicated by the red dot, which is smaller than the saddle point at $\chi_2\simeq (1+\sqrt{1-\gamma_2^2\lambda})/\gamma_2\simeq17$ and $V(\chi_1=0,\chi_2\simeq 17)/V_0\simeq38$ depicted by the yellow cross. The bottom panel shows the same trajectory in black, where each contour shows the height of the potential.}
    \label{fig: trajectory}
\end{figure}

Figure~\ref{fig: trajectory} shows the field trajectory, giving the height of the potential, as well.
The top panel is a projected three-dimensional plot of the trajectory, the red dot indicating the initial location of the fields.
The yellow cross marks the saddle point along the $\chi_2$ direction with $\chi_1\simeq 0$.
The bottom panel shows the same trajectory, where the contours represent the height of the potential.
In the present setup, $\chi_2$ may be identified as inflaton at early times and realizes the slow-roll inflation with the potential given by
\begin{align}
    V
    &\simeq
    V_0\chi_2^2 e^{-\gamma_2\chi_2}
\end{align}
for $\chi_1\simeq 0$.
Once the slow-roll regime is over, $\chi_2$ component alone starts oscillation and settles down at the local minimum where
\begin{align}
    \chi_1 
    &=
    {\gamma_1\lambda\over2h+\gamma_1^2\lambda},&
    \chi_2
    &=
    {\gamma_2\lambda\over2+\gamma_2^2\lambda}.
\end{align}
In the illustrated example, those values are $\chi_1\simeq 3$ and $\chi_2\simeq 0.006$.

\begin{figure}[htbp]
    \centering
    \includegraphics[width=.8\textwidth]{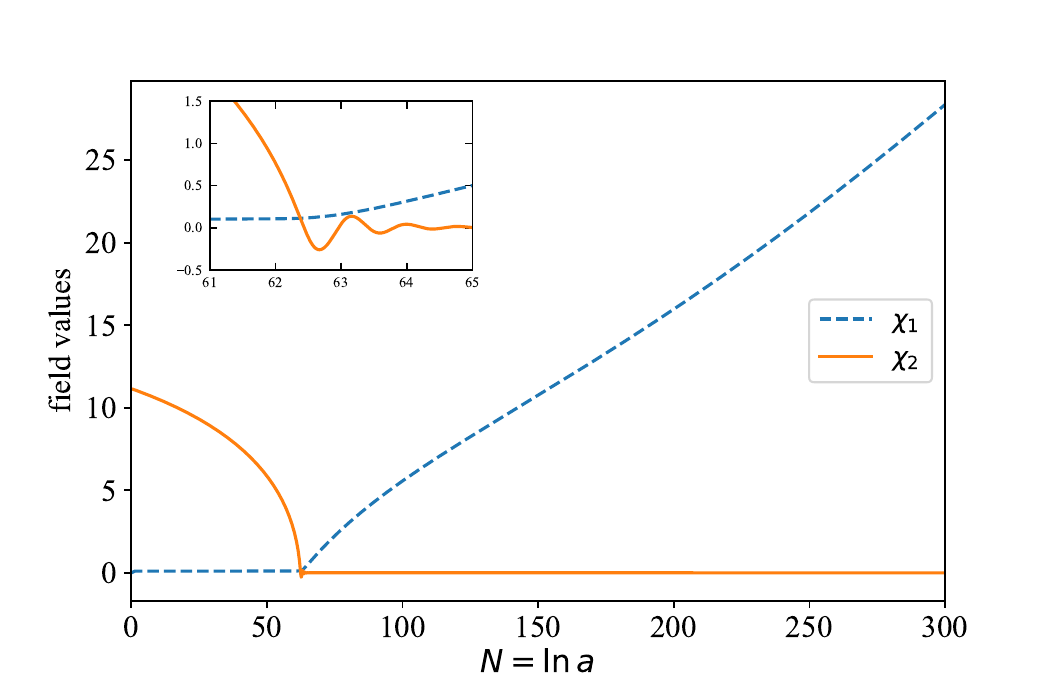}
    \caption{Shown is the evolution of each field with the same choice of the parameters as Fig.~\ref{fig: trajectory}, where the inset magnifies the oscillation period of $\chi_2$ in the same figure.}
    \label{fig: field evolution}
\end{figure}

Figure~\ref{fig: field evolution} shows the evolution of each field with the same choice of the parameters used in Fig.~\ref{fig: trajectory}. As is evident from the figure, $\chi_1$ is stabilized at $\chi_1\simeq 0$ during inflation due to the term $g\chi_1^2\chi_2^2$.
After the end of inflation, $\chi_2$ oscillation is damped, and some amount of its energy is transferred to $\chi_1$ through the same coupling, and $\chi_1$ is eventually expelled from the potential minimum toward $\chi_1\to\infty$, its effective potential being given by 
\begin{align}
    V(\chi_1\gg 1)
    &\simeq
    h V_0 \chi_1^2 e^{-\gamma_1\chi_1}.
\end{align}
This explains the dark energy at later times.
Notice that we may choose generic power $p$ on $\chi_1$ other than $p=2$, namely $V(\chi_1\gg1)\simeq
h V_0 \chi_1^p e^{-\gamma_1\chi_1}$, since the fixed point does not depend on $p$ as discussed in Sec.~\ref{sec: late time cosmology: dark energy}.
This power change may occur if the escape to field infinity is tilted along $\chi_2 = c \chi_1$, $c\neq 0$.

\begin{figure}[htbp]
    \centering
    \includegraphics[width=.8\textwidth]{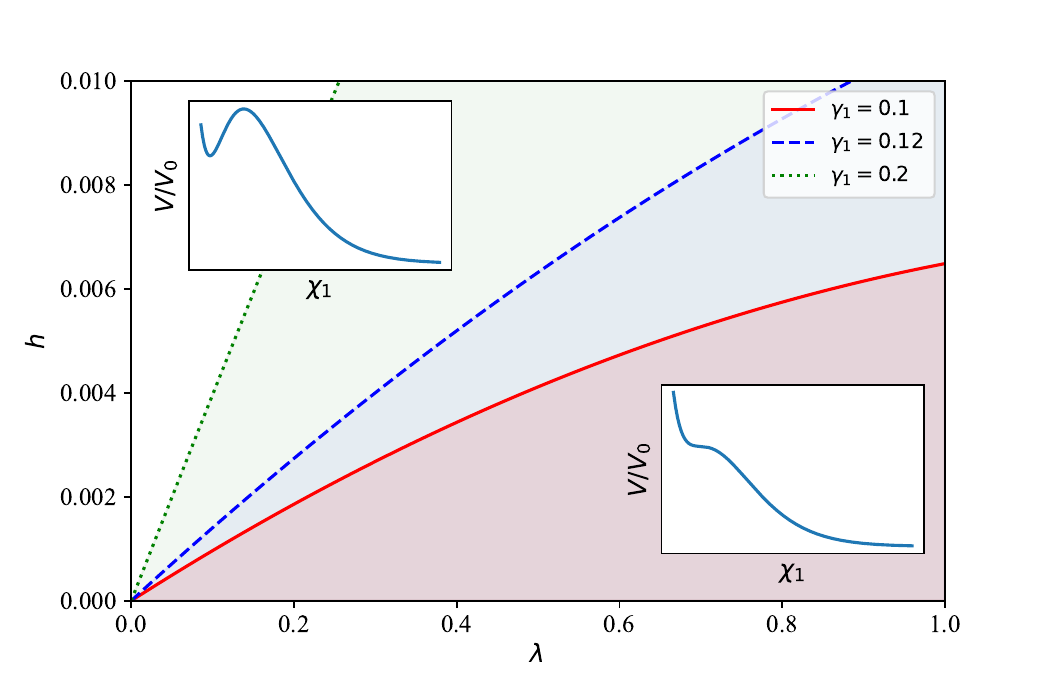}
    \caption{Each line shows $h_{\rm cr}$, below which the field excursion into infinity is possible, for a given $\gamma_1$. The field excursion toward $\chi_1\to\infty$ may be achieved in the shaded region below each line.}
    \label{fig: parameter space}
\end{figure}

It should be emphasized that the inflaton oscillation around the field origin is not only responsible for generating a heat bath but also for triggering the evolution of the dark energy field. Such behavior can be achieved for various choices of the potential parameters. Phenomenological constraints are imposed on $\gamma_1$ and $\gamma_2$. We fix $\gamma_2=0.12$ to satisfy the Planck data for inflation. Since observations indicate the equation-of-state parameter should be $w_\chi \sim -1$ in the present universe, we require $\gamma_1 < {\cal O}(0.1)$, as discussed in Sec.~\ref{sec: late time cosmology: dark energy}. Setting $g=1$ for simplicity, we study how the field trajectory depends on the choice of $\lambda$ and $h$ for a given $\gamma_1$. The initial condition is the same as in Fig.~\ref{fig: trajectory}. We find that, depending on the parameters, $\chi_1$ either rolls toward infinity or oscillates around the origin.

The boundary between the two cases is determined by the disappearance of the saddle point along the $\chi_1$ direction while fixing $\chi_2=\gamma_2\lambda/(2+\gamma_2^2\lambda) \equiv \chi_{2,{\rm min}}$. The condition for the disappearance can be obtained by finding the two extremum points along the $\chi_1$ direction of the potential $V(\chi_1, \chi_2=\chi_{2,{\rm min}})$. One of the two is a local minimum, and the other is a local maximum. Equating the two points, we obtain a relation among the potential parameters, namely, $g$, $\lambda$, $h$, $\gamma_1$, and $\gamma_2$. We define $h \equiv h_{\rm cr}$ satisfying this relation by
\begin{align}
    h_{\rm cr}(g,\lambda,\gamma_1,\gamma_2)
    &=
    {\lambda\over(2+\gamma_2^2\lambda)^2}
    \left[
        \gamma_1^2(1+\gamma_1^2\lambda)(4+\gamma_2^2\lambda)-g\gamma_2^2\lambda
    \right].
\end{align}
Here, we assume $h_{\rm cr} > 0$. The field excursion toward $\chi_1 \to \infty$ may be achieved when $h < h_{\rm cr}(g, \lambda, \gamma_1, \gamma_2)$. When $h > h_{\rm cr}$, conversely, $\chi_1$ oscillates around the potential minimum and eventually settles down. Figure~\ref{fig: parameter space} illustrates the boundary between these two cases. The red-solid, blue-dashed, and green-dotted lines represent $h_{\rm cr}$ as a function of $\lambda$ for $\gamma_1 = 0.1$, 0.12, and 0.2, respectively. In the shaded area below each line, the field excursion toward infinity is realized. The insets in the figure show the potential shape for $h < h_{\rm cr}$ and $h > h_{\rm cr}$. Notice that $h = 0$, equivalent to $p = 0$, is permissible, as the fixed point does not depend on $p$.

Before closing the section, we note that our numerical analysis does not include the radiation contribution in the total energy density, and thus the later time evolution of $\chi_1$ shown in Fig.~\ref{fig: trajectory} does not correspond to a realistic case.
However, as discussed in Sec.~\ref{sec: late time cosmology: dark energy}, the late time behavior has an asymptotic fixed point.
Therefore, we may expect that the late time evolution falls into such a case, though the detailed analysis on heating to a hot big bang universe is left for future study.

\section{Summary and discussion}
\label{sec: Summary and discussion}
In the context of eJBD gravity, we have proposed a phenomenological eJBD scalar potential of the power-times-exponential type, capable of explaining either inflation or dark energy, depending on the cosmological stage under consideration. The proposed potential is simple, involving only two parameters aside from the overall normalization of the energy scale: the power $p$ and the negative exponent $\gamma$. This potential features a local maximum at the field value $p/\gamma$, below which inflation occurs and above which dark energy is realized.

When the eJBD scalar is identified as the inflaton, the CMB data provides a constraint on the power, requiring $p<2$; otherwise, the model predicts an excessively large tensor-to-scalar ratio $r$ and an insufficiently small spectral tilt $n_s$. In this scenario, the initial field value of the inflaton is taken near the hilltop, and the inflaton slow-rolls toward the origin. For a viable choice of $p$, typically $\gamma$ cannot be greater than ${\cal O}(0.1)$, while too small a value for $\gamma$ is also not permitted. The parameter space for smaller $\gamma$ will be explored when reaching smaller $r$. Future CMB observations, including LiteBIRD~\cite{LiteBIRD:2022cnt}, CMB-S4~\cite{CMB-S4:2016ple}, and SO~\cite{SimonsObservatory:2018koc}, are capable of testing $r$ at $10^{-3}$ level, which can potentially probe our scenario.

The eJBD scalar may also be identified as the quintessence field that accounts for dark energy at later times. We have shown that there exists an asymptotic fixed point where the universe is fully dominated by $\chi$, and the equation-of-state parameter is determined solely by $\gamma$, specifically $w_\chi = -1 + \gamma^2/3$. Since the present stage of the universe has not yet been fully dominated by the quintessence field, $\chi$ is on its way to reaching field infinity. A deviation of $w_\chi$ from $-1$ should, in principle, be testable by high-precision observations. Up-to-date data from DESI~\cite{DESI:2024mwx} result in $w_\chi(a_0)=-0.99^{+0.15}_{-0.13}$ (68\%CL), indicating $\gamma\lesssim 0.7$, which will be further improved, including its time dependence, in future dark energy survey, e.g., Euclid~\cite{Euclid:2024yrr} and LSST~\cite{2009arXiv0912.0201L}.

Given the successful cosmology at early and late times, we have proposed a unified model to smoothly connect the regimes of inflation and dark energy. Our numerical analysis has shown that inflation concludes with a period of oscillation, which amplifies the other field direction and eventually expels it from the potential origin. The field excursion of quintessence at later times is governed by the potential of the power-times-exponential form. 

Our numerical analysis does not account for radiation production, necessitating further study on how the rolling field dissipates into the produced particles that realize a heat bath. A detailed analysis of the heat-up stage, including the effects of parametric amplification, is left for future work.

\section*{Acknowledgements}
This research was partially supported by the JSPS KAKENHI Grant Nos.~21H01107 (KO) and 21K03575 (MY) from the Japanese Ministry of Education, Culture, Sports, Science, and Technology.

\appendix

\section{Autonomous system for a generic potential}
\label{appendix: Fixed points}

We introduce a generic parametrization to efficiently derive, if any, a closed set of differential equations.
An important difference compared to the literature, e.g. Refs.~\cite{Copeland:1997et,Copeland:2006wr}, is that we take the equation-of-state parameter $w$ to be also a function of time, since it is determined by the dynamics of the system.

For a scalar field $\chi$ with a potential $V(\chi)$, we define the following quantities:
\begin{align}
    x\equiv \frac{\dot \chi}{\sqrt{6}H},\;\;\;
    y\equiv \frac{\sqrt{V}}{\sqrt{3}H},\;\;\;
    z_1\equiv \frac{\partial_\chi V}{V},\;\;\;
    z_2\equiv \frac{{\partial^2_\chi V}}{\partial_\chi V},\;\;\;
    z_3\equiv \frac{{\partial^3_\chi V}}{\partial^2_\chi V},
    \cdots.\label{eq: x, y, z}
\end{align}
By assuming that $\chi$ is secluded from other sectors, the equation of motion is given by
\begin{align}
    \ddot\chi + 3H\dot\chi + \partial_\chi V(\chi) = 0.\label{eq: EoM for chi}
\end{align}
The equation of state parameter $w$ is introduced as
\begin{align}
    w = \frac{p_{\rm tot}}{\rho_{\rm tot}}
\end{align}
for the total pressure $p_{\rm tot}$ and the total energy density $\rho_{\rm tot}$, so that we may write $\dot H$ as
\begin{align}
    \dot H &= -\frac{1}{2}\sum_i(\rho_i+p_i)=-\frac{3}{2}H^2(1+w),\label{eq: H-dot 1}
\end{align}
where the sum over $i$ is taken for $i=r$ (radiation), $i=m$ (matter), and $i=\chi$.
Since $p_r/\rho_r=1/3$, $p_m/\rho_m=0$, and $\rho_\chi+p_\chi=\dot\chi^2$, $\dot H$ may also be written as
\begin{align}
    \dot H &= -\frac{1}{2}\left(
        \frac{4}{3}\rho_r+\rho_m + \dot\chi^2
    \right).\label{eq: H-dot 2}
\end{align}
From Eqs.~(\ref{eq: H-dot 1}) and (\ref{eq: H-dot 2}), we obtain
\begin{align}
    w &= \frac{\rho_r}{9H^2}+x^2-y^2, \label{eq: w}
\end{align}
where we have used $\rho_r+\rho_m=3H^2(1-x^2-y^2)$.
Using Eqs.~(\ref{eq: x, y, z}), (\ref{eq: H-dot 1}), and (\ref{eq: w}), we obtain
\begin{align}
    x' &= \frac{3}{2}(w-1)x-\sqrt{\frac{3}{2}}y^2z_1,\\
    y' &= \frac{3}{2}(w+1)y+\sqrt{\frac{3}{2}}xyz_1,\\
    w' &= 3w\left(w-\frac{1}{3}\right)-2x^2-4y^2-2\sqrt{6}xy^2z_1,\\
    z'_i &= \sqrt{6}x(z_iz_{i+1}-z_i^2),
\end{align}
where the prime denotes $d/dN$ for $N=\ln a$, and $z'_i$ ($i=1,2,3,\cdots$) for a sufficiently large $i$ should be considered until the system is closed, namely, the right-hand side of $z'_i$ does not include $z_{i+1}$ when expressing $z_{i+1}$ in terms of $z_1, \cdots, z_i$.
Note that the initial condition for $w$ should be taken so that $w=(1+2x^2-4y^2-\Omega_m)/3$ where we may assume $\Omega_m=0$ at early times.

We turn to apply the analysis explained above to the specific case where the potential is given by
\begin{align}
    V(\chi)
    &=
    V_0\chi^p e^{-\gamma\chi},
\end{align}
assuming $\chi>0$.
The autonomous system may be analyzed using a set of first order differential equations,
\begin{align}
    x' &= \frac{3}{2}(w-1)x-\sqrt{\frac{3}{2}}y^2z_1,\\
    y' &= \frac{3}{2}(w+1)y+\sqrt{\frac{3}{2}}xyz_1,\\
    w' &= 3w\left(w-\frac{1}{3}\right)-2x^2-4y^2-2\sqrt{6}xy^2z_1,\\
    z'_1 &= -\frac{(z_1+\gamma)^2}{p}.
\end{align}
Taking $x'=0$, $y'=0$, $w'=0$, and $z'_1=0$, we find that
\begin{align}
    x &= \frac{\gamma}{\sqrt6},&
    y &= \sqrt{1-\frac{\gamma^2}{6}},&
    w_\chi &= -1 + \frac{\gamma^2}{3}
\end{align}
is the only viable fixed point where the solution does not depend on $p$.
Note that the asymptotic value of the density parameter is $\Omega_\chi(t\to\infty)=x^2+y^2=1$, and thus the present universe is located midway towards the fixed point, as discussed in Sec.~\ref{sec: late time cosmology: dark energy}.

The stability of the fixed point can be analyzed by perturbing $x$, $y$, $z_1$, and $w$ about the fixed point.
Denoting $\vec r=(x,y,z_1,w)^T$, we consider $\vec r=\vec r_0 + \delta \vec r$ where $\vec r_0$ represents the fixed point. 
The coupled differential equations can then be recast into the form of
\begin{align}
    \delta \vec r \,' &= A \delta \vec r,
\end{align}
where $A$ is a $4\times4$ matrix given by
\begin{align}
    A
    &=
    \begin{pmatrix}
        {-6+\gamma^2\over 2} & \gamma\sqrt{6-\gamma^2} & {-6+\gamma^2\over 2\sqrt6} & {1\over2}\sqrt{3\over2}\gamma \\
        -{\gamma\sqrt{6-\gamma^2}\over2} & 0 & {\gamma\over2}\sqrt{1-{\gamma^2\over 6}} & {1\over2}\sqrt{9-{3\gamma\over2}}\\
        0 & 0 & 0 & 0\\
        -\sqrt{2\over3}\gamma(-4+\gamma^2) & 2\sqrt{4-{2\gamma^2\over 3}}(-2+\gamma^2) & {\gamma\over 3}(-6+\gamma^2) & -7+2\gamma^2
    \end{pmatrix},
\end{align}
whose eigenvalues are $\{0,(-6+\gamma^2)/2,-4+\gamma^2,-3+\gamma^2\}$.
Therefore, the fixed point is an attractor when $\gamma<\sqrt{3}$ as $\delta \vec r\to0$ for $t\to\infty$.

\section{Slow-roll conditions and the end of inflation}
\label{appendix: Slow-roll conditions and the end of inflation}

The field value at the end of inflation, $\chi_{\rm end}$, can be found by imposing either $\epsilon(\chi_{\rm end})=1$ or $\eta(\chi_{\rm end})=\pm1$ for $0<\chi_{\rm end}<\chi_{\rm max}$.
To find $\chi_{\rm end}$ in a simple manner, we introduce a parameter $x$ such that
\begin{align}
    \chi 
    &=
    \frac{\chi_{\rm max}}{1+x}
\end{align}
in the region $0<x<\infty$.
The slow-roll parameters may then be written as
\begin{align}
    \epsilon(x)
    &=
    \frac{\gamma^2}{2}x^2,
    &
    \eta(x)
    &=
    \gamma^2x^2-\frac{\gamma^2}{p}(x+1)^2.
\end{align}
For $p>2$, $\eta$ quickly becomes positive for $x>0$, and $\epsilon=\eta$ is achieved at
\begin{align}
    x
    &=
    \frac{1}{\sqrt{p/2}-1}
    \equiv x_*,
\end{align}
where $\epsilon=\eta=\left(\frac{\gamma}{\sqrt{p}-\sqrt{2}}\right)^2$.
We find $\epsilon > \eta$ for $x<x_*$, while $\epsilon <\eta$ for $x>x_*$, which indicates that for $\gamma>\sqrt{p}-\sqrt{2}$, $\chi_{\rm end}$ is determined by $\epsilon(\chi_{\rm end})=1$, and for $\gamma<\sqrt{p}-\sqrt{2}$, $\eta(\chi_{\rm end})=1$.
Therefore, we obtain
\begin{align}
    \chi_{\rm end}
    &=
    \begin{cases}
        \frac{p}{\gamma+\sqrt{2}} & (p>2, \gamma>\sqrt{p}-\sqrt{2})\\
        \frac{p-1}{2} & (p>2,\gamma=1)\\
        \frac{p\gamma-\sqrt{p(\gamma^2+p-1)}}{\gamma^2-1} & (p>2,\gamma\neq1,\gamma<\sqrt{p}-\sqrt{2})
    \end{cases}.
\end{align}
For $p=2$, given that $\gamma^2\ll p$, $\epsilon>|\eta|$ is always satisfied, and thus $\chi_{\rm end}$ is determined by $\epsilon(\chi_{\rm end})=1$.
For $p\leq2$, $\chi_{\rm end}$ can be found in the same manner, and here we only show the case with $p=1$ and $2$ as examples:
\begin{align}
    \chi_{\rm end}
    &=
    \begin{cases}
        \frac{2}{\gamma+\sqrt{2}} & (p=2,\gamma<2)\\
        \frac{1}{\gamma+\sqrt{2}} & (p=1,\gamma<1)
    \end{cases}.
\end{align}

\section{Linear-times-exponential potential}
\label{appendix: Linear-times-exponential potential}

A possible origin of the potential for an eJBD field $\chi$ is a quantum effect.
For instance, we may consider a Majorana fermion whose mass is given by a function of $\chi$, $M(\chi)$.
By integrating it out, we are left with an effective potential
\begin{align}
    V_{\rm 1-loop}(\chi)
    &=
    -\frac{M^4(\chi)}{32\pi^2}\ln\frac{M^2(\chi)}{\mu^2},
\end{align}
where $\mu$ is the renormalization scale.
If we suppose
\begin{align}
    M(\chi)
    &=
    \mu e^{-\gamma\chi/4},
\end{align}
the potential becomes
\begin{align}
    V_{\rm 1-loop}(\chi)
    &=
    \frac{\gamma\mu^4}{64\pi^2}\chi e^{-\gamma\chi},
\end{align}
where we may identify $V_0=\gamma\mu^4/64\pi^2$.
Note that this potential is unstable for $\chi<0$, which is eliminated by, for instance, replacing $\chi$ by $|\chi|$.
The region where $\chi>0$ is nevertheless relevant to explain the dark energy.

\nocite{*}
\bibliographystyle{JHEP}
\bibliography{biblio}

\end{document}